\documentclass[conference]{IEEEtran}
\usepackage{graphicx}
\usepackage{latexsym}
\usepackage{amsmath}
\usepackage{url}
\begin{document}
\title{Reasoning about Identifier Spaces:\\
How to Make Chord Correct}
\author{\IEEEauthorblockN{Pamela Zave}
\IEEEauthorblockA{AT\&T Laboratories---Research\\
Bedminster, New Jersey 07921, USA\\
Email: pamela@research.att.com}}
\maketitle

\begin{abstract}
The Chord distributed hash table (DHT) is well-known and often used to
implement peer-to-peer systems.
Chord peers find other peers, and access their data,
through a ring-shaped pointer
structure in a large identifier space.
Despite claims of proven correctness, i.e., eventual reachability,
previous work
has shown that the Chord ring-maintenance protocol is not correct
under its original operating assumptions.
Previous work
has not, however, discovered whether Chord could be made correct
under the same assumptions.
The contribution of this paper is to provide the first specification
of correct operations and initialization
for Chord, an inductive invariant that is necessary and
sufficient to support a proof of correctness, and two independent
proofs of correctness.
One proof is informal and intuitive, and applies to networks of any size.
The other proof is based on a formal model in Alloy, and uses
fully automated analysis to prove the assertions for networks of
bounded size.
The two proofs complement each other in several important ways.
\end{abstract}

\section{Introduction}
\label{sec:intro}

Peer-to-peer systems are distributed systems featuring decentralized
control, self-organization of similar nodes, fault-tolerance,
and scalability.
The best known peer-to-peer system is Chord, which
was first presented in a 2001 
SIGCOMM paper \cite{chord-sigcomm}.
This paper was the fourth-most-cited paper in computer science for
several years (according to Citeseer), and won
the 2011 SIGCOMM Test-of-Time Award.

The Chord protocol maintains a network of nodes that can reach
each other despite the fact that autonomous
nodes can join the network, leave
the network, or fail at any time.
The nodes of a Chord network have identifiers in an {\it m}-bit
identifier space, and
reach each other through pointers in this identifier space.
Because the network structure is based on adjacency in the identifier
space, and $2^{m} - 1$ is adjacent to 0, the structure of a
Chord network is a ring.

A Chord network is usually used to maintain a distributed hash table (DHT),
which is a key-value store in which the keys are also identifiers
in the same {\it m}-bit space.
In turn, the hash table can be used to implement
shared file storage, group directories, and many
other purposes.
Chord has been implemented many times, and used to build large-scale
applications such as BitTorrent.
And the continuing influence of Chord is easy to trace in more recent
systems such as Dynamo \cite{dynamo}.

The basic correctness property for Chord is eventual
reachability:  given ample time and no further joins, departures, or
failures,
the protocol can repair all defects in the ring structure.
If the protocol is not correct in this sense, then some nodes of a
Chord network will become permanently unreachable from other nodes.
The introductions of the original Chord papers
\cite{chord-sigcomm,chord-ton}
say, ``Three features that
distinguish Chord from many other peer-to-peer lookup protocols
are its simplicity, provable correctness, and provable performance.''
An accompanying PODC paper \cite{chord-podc}
lists invariants of the ring-maintenance protocol.

The claims of simplicity and performance are certainly true.
The Chord algorithms are far simpler and more completely specified
than those of other DHTs, such as Pastry \cite{pastry},
Tapestry \cite{tapestry}, CAN \cite{CAN}, and Kademlia \cite{kademlia}.
Operations are fast because
there are no atomic operations requiring locking of multiple nodes,
and even queries are minimized.

Unfortunately, the claim of correctness is not true.
The original specification with its original operating
assumptions does not have eventual reachability,
and {\it not one} of the seven properties claimed to be invariants
in \cite{chord-podc} is actually an invariant
\cite{chord-ccr}.
This was revealed by modeling the protocol in the Alloy language
and checking its properties with the Alloy Analyzer \cite{alloy-book}.

The principal contribution of this paper
is to provide the first specification of a
version of Chord that is as efficient as the original,
correct under reasonable operating
assumptions, and actually proved correct.
The new version
corrects all the flaws that were revealed in \cite{chord-ccr},
as well as some additional ones.
The proof provides a great deal of insight into how rings in identifier
spaces work, and is backed up by a formal, analyzable model.

Some motivations and possible benefits of this work are presented
below.
They are categorized according to the audience or constituency that
would benefit.

{\it For those who implement Chord or rely on a Chord implementation:}
It seems obvious that implementers
should have a precise and correct specification
to follow.
They should understand the operating assumptions so as not to undermine
them.
They should also know the invariant for Chord, as dynamic checking
of the invariant is a
design principle for enhancing security in distributed
systems \cite{sitmorris}.

Critics of this work have claimed 
that all the flaws in original Chord are
either obvious and fixed by all implementers,
or extremely unlikely to cause trouble during Chord execution.
It is a fact that some implementations retain original flaws,
citing \cite{overlog} not because it is a bad implementation, but simply
because the code is published and readable.
Concerning whether the flaws cause real trouble or not, 
Chord implementations are 
certainly reported to have been unreliable.
It is in the nature of distributed systems that failures are difficult
to diagnose, and no one knows (or at least tells) what is really going on.
Any means for increasing the reliability of distributed systems,
especially without sacrificing efficiency, is an unmixed blessing.

{\it For those interested in building more robust or more functional
peer-to-peer systems based on Chord:}
Due to its simplicity and efficiency, it is an attractive idea to
extend original Chord with stronger guarantees and additional
properties. 
Work has already been done on
protection against malicious peers
\cite{awerbuch-robust,chord-byz,sechord},
key consistency and data consistency \cite{scatter},
range queries \cite{rangequeries},
and atomic access to replicated data \cite{atomicchord,etna}.

For those who build on Chord, and reason about Chord behavior,
their reasoning should have a sound foundation. 
Previous
research on augmenting and strengthening Chord, as referenced above,
relies on ambiguous descriptions of Chord and unsubstantiated claims about
its behavior.
These circumstances can lead to misunderstandings about how Chord
works, as well as to unsound reasoning.
For example,
the performance analysis in \cite{chord-churn} makes the assumption
that every operation of a particular kind makes progress according to
a particular measure, which is easily seen to be false \cite{chord-ccr}.

{\it For those interested in encouraging application of formal methods:}
This project has already had an impact,
as developers at Amazon
credit the discovery of Chord flaws \cite{chord-ccr} with convincing
them that formal methods can be applied productively to real distributed
systems \cite{amazon}.

The proof of correctness is also turning out to be an important case
study.
In this paper there are two independent proofs, one informal and
one by model checking.
The informal proof applies to networks of any size, and
provides deep insight into how and why the
protocol works.
The Alloy
model with its automated checking applies only to networks of bounded
size, and offers limited insight, but it is an
indispensable backup to the informal proof because it guards against
human error.
Also, it was an indispensable precursor to finding the general proof,
because it indicated which theorems were likely to be true.

For those interested in formal proofs,
the Alloy-only proof in \cite{chord-arxiv} has been used as a test case for
the Ivy proof system \cite{ivy}, and the new proof given here is being
used as a test case for the Verdi proof system \cite{verdi}.

Finally, there are other possible uses for ring-shaped
pointer structures in large identifier spaces 
({\it e.g.,} \cite{awerbuch-hyperring,CAN}).
The reasoning about identifier spaces used in this paper may also be
relevant to other work of this kind.

The paper begins with an overview of Chord using the revised, correct
ring-maintenance operations
(Section~\ref{sec:overview}), and a specification of these new
operations (Section~\ref{sec:spec}).
Although the specification
is pseudocode for immediate accessibility, it is a paraphrase
of the formal model in Alloy.

Correct operations are necessary but not sufficient.
It is also necessary to initialize a network correctly.
Original Chord is initialized with a network of one node,
which is not correct, and
Section~\ref{sec:initialization} shows why.
This section also introduces the inductive invariant for the proof,
because a Chord network can safely be initialized in any state that
satisfies the invariant.

Summarizing the previous two sections,
Section~\ref{sec:diff} compares the revised Chord protocol
with the original version, explaining how they differ.
Together Sections~\ref{sec:initialization} and
\ref{sec:diff} present most of the problems with original
Chord reported
in \cite{chord-ccr} (as well as previously unreported ones).
The problems are not presented first because they make more sense when
explained along with their underlying nature and how to remove them.

The proof of correctness is largely based on reasoning about ring
structures in identifier spaces.
Section~\ref{sec:idspace} presents some useful theorems about these
spaces and shows how they apply to Chord.
The actual proof in
Section~\ref{sec:proof} follows a fairly conventional outline.
Section~\ref{sec:alloy} discusses the formal model and 
model-checked version of the proof, while Section~\ref{sec:related}
covers related and future work.

\begin{figure}
\centering
\includegraphics[scale=0.80]{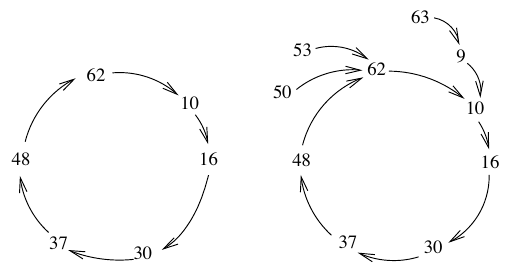}
\caption{Ideal (left) and valid (right) networks.
Members are represented by their identifiers.
Solid arrows are successor pointers.}
\label{fig:valid}
\end{figure}

\section{Overview of correct Chord}
\label{sec:overview}

Every member of a Chord network has an identifier (assumed unique) that
is an {\it m}-bit hash of its IP address.
Every member has a {\it successor list} of pointers to other members.
The first element of this list is the {\it successor}, and is
always shown as a solid
arrow in the figures.
Figure~\ref{fig:valid} shows two Chord networks with {\it m} = 6,
one in the ideal state of a ring ordered by identifiers,
and the other
in the valid state of an ordered ring with appendages.
In the networks of Figure~\ref{fig:valid}, key-value pairs with keys
from 31 through 37 are stored in member 37.
While running the ring-maintenance protocol, a member also acquires and
updates a {\it predecessor} pointer, which is always shown as a dotted
arrow in the figures.

\begin{figure*}
\centering
\includegraphics[scale=0.80]{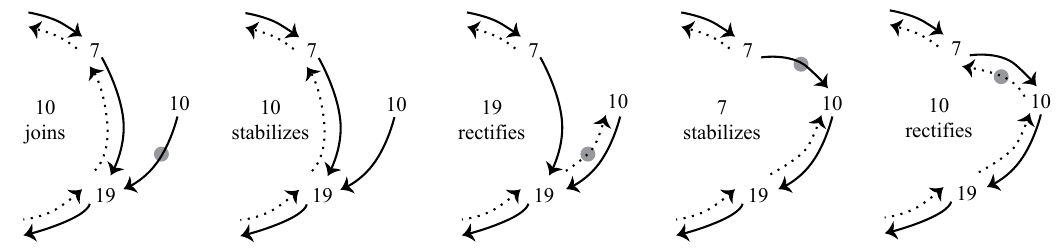}
\caption{A new node becomes part of the ring.
A gray circle marks the pointer updated by an operation, if any.
Dotted arrows are predecessors.}
\label{fig:join}
\end{figure*}

The ring-maintenance protocol is specified in terms of four
operations, {\it join, stabilize, rectify,} and {\it fail.}
Each operation
is executed by a member and changes only the state of that member.
In executing an operation, the member often queries another member or
sequence of members, 
updating its own pointers as necessary after getting
the result of each query.
The specification of Chord assumes that inter-node communication is
bidirectional and
reliable, so we are not concerned with Chord behavior when inter-node
communication fails.

A node becomes a member in a {\it join} operation.
A member is also referred to as a {\it live} node.
When a member joins, it contacts some existing member to look up
a member that is near to it in identifier space, and gets a
successor list from that nearby member.
The first stage of Figure~\ref{fig:join} shows successor and predecessor
pointers in a section of a network where 10 has just joined.

When a member {\it stabilizes}, it learns its successor's predecessor.
It adopts the predecessor
as its new successor, provided that the predecessor
is closer in identifier order
than its current successor.
Because a member must query its successor to stabilize, this is also
an opportunity for it to update its successor list with information
from the successor.
Members schedule their own stabilize operations, which should
be periodic.

Between the first and second stages
of Figure~\ref{fig:join}, 10 stabilizes.
Because its successor's predecessor is 7, which is not a better successor
for 10 than its current 19, this operation does not change the successor
of 10.

After stabilizing (regardless of the result),
a node notifies its successor of its identity.
This causes the notified member to execute a {\it rectify} operation.
The rectifying member
adopts the notifying member as its new predecessor
if the notifying member
is closer in identifier order than its current predecessor,
or if its current predecessor is dead.
In the third stage of Figure~\ref{fig:join},
10 has notified 19, and 19 has adopted 10 as its new predecessor.

In the fourth stage of Figure~\ref{fig:join}, 7 stabilizes, which causes
it to adopt 10 as its new successor.
In the last stage 7 notifies and 10 rectifies, 
so the predecessor of 10 becomes 7.
Now the new member 10 is completely incorporated into the ring, and all
the pointers shown are correct.

The protocol requires that a member or live node always
responds to queries in a timely fashion.
A node ceases to be a member in a {\it fail} operation, which can
represent failure of the machine, or the node's
silently leaving the network.
A member that has failed is also referred to as a {\it dead} node.
The protocol also requires that, after a member fails, it no longer
responds to queries from other members.
With this behavior, members can detect the failure of other members
perfectly by observing whether they respond to a query before a timeout
occurs.
Failed nodes can rejoin later by executing a new {\it join} operation.

Failures can produce gaps in the ring, which are repaired during
stabilization.
As a member attempts to query its successor for stabilization, 
it may find that its
successor is dead.
In this case it attempts to query the next member in its successor
list and make this its new successor, continuing through the list
until it finds a
live successor.

Chord relies on a critical operating assumption that
a member always has a live successor in its list.
In practice, an implementer must maintain the truth of this assumption
by adjusting the length of successor lists (a parameter
of the algorithm) and the rate of stabilization (which is not formally
specified) to compensate for the failure rate.
If successor lists are long enough to provide adequate redundancy,
and stabilization occurs often enough to replace dead successors with
live ones quickly, then a successor list should
always have at least one live
member remaining, even after an entry in its list fails.

As in the original Chord papers \cite{chord-sigcomm,chord-ton},
we wish to define a correctness property of eventual reachability:
given ample time and no further disruptions, 
the ring-maintenance protocol can repair defects so that
every member of a Chord network is reachable from every other member.
Note that a network with appendages (nodes 50, 53, 63, 9 on the right
side of Figure~\ref{fig:valid}) cannot have full reachability,
because an appendage cannot be reached by a member that is not in
the same appendage.
The correctness property here is slightly stronger, being based
on the definition of the property {\it Ideal.}

\vspace{.1 in}
{\it Definition:} A Chord network is in the {\it Ideal} state if:
\begin{itemize}
\item
every pointer points to a live node;
\item
every successor pointer is its node's
first successor in identifier order;
\item
every predecessor
pointer is its node's first predecessor in identifier
order; and 
\item
the tail of the successor list of a node (past its head entry or
successor) is the successor's successor list,
with the last entry removed.
\end{itemize}
\vspace{.1 in}
For example, on the right of Figure~\ref{fig:valid}, the ideal
successor of 48 is 50 because 50 is the closest successor to
48 in identifier order.
The correctness property we need to prove is:

\vspace{.1 in}
{\it Starting in any execution state, if there are no
subsequent join or fail operations, then eventually the network will become
Ideal and remain Ideal.}

\vspace{.1 in}
Defining a member's {\it best successor} as its the first
entry in its successor list pointing
to a live node, a {\it ring member} is a member that can reach
itself by following the chain of best successors.
An {\it appendage member} is a member that is not a ring member.
Of the seven invariants presented in
\cite{chord-podc} (and all violated by original Chord),
the following four are necessary for correctness.
\begin{itemize}
\item
There must be a ring, which means that there must be a non-empty set
of ring members ({\it AtLeastOneRing}).
\item
There must be no more than one ring, which means that from each ring
member, every other ring member is reachable by following the chain
of best successors ({\it AtMostOneRing}).
\item
On the unique ring, the nodes must be in identifier 
order ({\it OrderedRing}).
\item
From each appendage member, the ring must be reachable by following
the chain of
best successors ({\it ConnectedAppendages}).
\end{itemize}
If any of these properties is violated,
there is a defect in the structure that
the ring-maintenance protocol cannot repair.
If there is no ring, then ring-based reachability will not work.
If there is more than one ring, then the network has separated into
disjoint subnetworks.
If appendage members cannot reach the ring, then the appendage is
a disconnected fragment.
A network that violates {\it OrderedRing} does not seem as
catastrophic, but it impedes lookup, and 
the protocol cannot fix it \cite{chord-podc}.
It follows that any inductive invariant must imply these properties.

The Chord papers define the lookup protocol, 
which is used to find the member primarily responsible for a key,
namely the ring member
with the smallest identifier greater than or equal to the key.
The lookup protocol is not discussed further here.
Chord papers also define the maintenance and use of finger tables, which 
greatly improve lookup speed
by providing pointers that cross the ring like chords of a circle.
Because finger tables are an optimization and
they are built from successor lists, their correctness relies on
the correctness of successor lists.
Finger tables are not discussed further here.

\section{Specification of ring-maintenance operations}
\label{sec:spec}

In this section, the operations, data structures, and related material
are presented in pseudocode.
Although pseudocode is not analyzable as the Alloy model is, it
translates more directly to implementation code.

\subsection{Identifiers and node state}
\label{sec:state}

There is a type 
{\it Identifier}
which is a string of {\it m} bits.
Implicitly, whenever a member transmits the identifier of a member, it
also transmits its IP address so that the recipient can reach the 
identified member.
The pair is self-authenticating, as the identifier must be the hash of the
IP address according to a chosen function.

The Boolean function 
{\it between}
is used to test the order of
identifiers.
Because identifier order wraps around at zero,
it is meaningless to test the order of two identifiers---each precedes
and succeeds the other.
This is why 
{\it between}
has three arguments:
\small
\begin{verbatim}
Boolean function between (n1,nb,n2: Identifier) 
{  if (n1 < n2) return ( n1 < nb && nb < n2 )
   else         return ( n1 < nb || nb < n2 )     
}
\end{verbatim}
\normalsize
For 
{\it nb}
to be 
{\it between n1}
and 
{\it n2},
it must be equal to neither.
Further properties of identifier spaces
are presented in Section~\ref{sec:idspace}.

Each node that is a member of a Chord network has the following 
state variables.
They are all initialized by the {\it join} operation.
\small
\begin{verbatim}
myIdent: Identifier;
prdc: Identifier;
succList: list Identifier;      // length is r
\end{verbatim}
\normalsize
Note that {\it myIdent}
is the hash of its IP address, and
{\it prdc} is the node's predecessor.
{\it succList} is the node's entire successor list; the head of this list
is its {\it successor}.
The parameter {\it r} is the fixed length of all successor lists.

\subsection{Maintaining a shared-state abstraction}
\label{sec:shared}

Reasoning about Chord requires reasoning about the global state,
so the protocol must maintain the abstraction of a shared,
global state.
To do this, the algorithmic steps of the protocol must behave as if
atomic and interleaved.
In each algorithmic step, a node reads the state of at most one other
node,
and modifies only its own state.

In an implementation, a node reads the state of another node by
querying it.
If the node does not respond within a time parameter {\it t}, then it
is presumed dead.
If the node does respond, then the atomic step associated with the query
is deemed to occur at the instant that the queried node responds with
information about its own state.

To maintain the shared-state abstraction, the querying node
must obey the following rules:
\begin{itemize}
\item
The querying node does not know the instant that its query is
answered; it only knows that the response was sent
some time after it sent the query.
So the querying node must treat its own state, between the time it
sends the query and the time it finishes the step by updating
its own state, as undefined.
The querying node cannot respond to queries about its state
from other nodes during this time.
\item
If the querying node is delaying response to a query because it is
waiting for a response to its own query, it must return interim
``response pending'' messages so that it is not presumed dead.
\item
If a querying node is waiting for a response, and is queried by another
node just to find out if it is alive or dead, it can respond immediately.
This is possible because the response does not contain any information
about its state.
\end{itemize}
This covers all possibilities except that of a deadlock due to circular
waiting for query responses.
Freedom from deadlock is covered in the proof of correctness in 
Section~\ref{sec:proof}.

\subsection{Join and fail operations}
\label{sec:joinfail}

When a node is not a member of a Chord network, it has no Chord state
variables, and does not respond to queries from Chord members.
To join a Chord network, a node must first calculate its own
Chord identifier 
{\it myIdent}.
It must also know some member of the network---it does not even
matter whether it is a ring member or appendage---and 
must ask the member to use the
lookup protocol to find a member 
{\it newPrdc}
such that
{\it between (newPrdc, myIdent, head(newPrdc.succList))}. 

Provided with this information, the node joins in a single atomic
step, by executing the following pseudocode:
\small
\begin{verbatim}
// Join step

// newPrdc has value from previous lookup
newPrdc: Identifier;

query newPrdc for newPrdc.succList;
if (query returns before timeout) {
   succList = newPrdc.succList;
   prdc = newPrdc;
}
else abort;
\end{verbatim}
\normalsize
If the query fails then
{\it newPrdc}
has died, and the node has no
choice but to try joining again later.

A fail operation is also a single atomic step.
When a member node fails or leaves a Chord network,
it deletes its Chord state variables and ceases to respond to queries.
Fortunately, the proof of correctness shows that a node can re-join
safely even if other nodes still have pointers to it from its
former episode of membership.

\subsection{Stabilize and rectify operations}
\label{sec:stabilize}

A stabilize operation may require a sequence of steps.
First, the stabilizing node executes a {\it StabilizeFromSuccessor}
step:
\small
\begin{verbatim}
// StabilizeFromSuccessor step 

// newSucc not initialized
newSucc: Identifier;

query head(succList) for 
      head(succList).prdc and
      head(succList).succList;
if (query returns before timeout) {
   // successor live, adopt its list as mine
   succList = 
      append ( 
         head(succList),
         butLast(head(succList).succList)
      );
   newSucc = head(succList).prdc;
   if (between(myIdent,newSucc,head(succList))) 
      // predecessor may be a better successor
      next step is StabilizeFromPredecessor;
      // else stabilization is complete
}
// successor is dead, remove from succList
else 
   succList = 
      append(tail(succList),last(succList)+1);
   next step is StabilizeFromSuccessor again;
\end{verbatim}
\normalsize

First the
node queries its successor for its successor's predecessor and
successor list.
If this query times out, then the node's successor is presumed dead.
The node removes the dead successor from its successor list and 
does another {\it StabilizeFromSuccessor} step.\footnote{The empty
place in the successor list is filled with an artificial entry at the
end, created by adding one to the last real entry.
The reason for this entry will be made clear by the proof.}
We know that eventually it will find a live successor in its list,
because of the operating assumption (from Section~\ref{sec:overview})
that successor lists are long enough so that each list contains at
least one live node.

Once the node has contacted a live successor, 
it adopts its successor list (all but the last entry) as its own
second and later successors.
It then tests the successor's predecessor to see if it might be a
better first successor.
If so, the node then executes a {\it StabilizeFromPredecessor} step.
If not, the stabilization operation is complete.

\begin{figure*}
\centering
\includegraphics[scale=0.80]{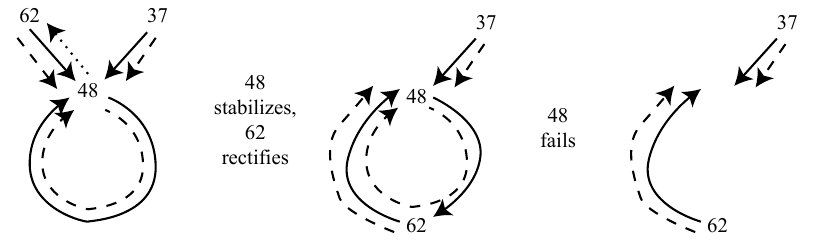}
\caption{Why the ring cannot be initialized at size 1.
Dashed arrows are second-successor pointers.
Predecessor pointers are not shown in the last two stages, as they are
irrelevant.
This problem was not reported in \cite{chord-ccr}.}
\label{fig:init}
\end{figure*}

\begin{figure*}
\centering
\includegraphics[scale=0.80]{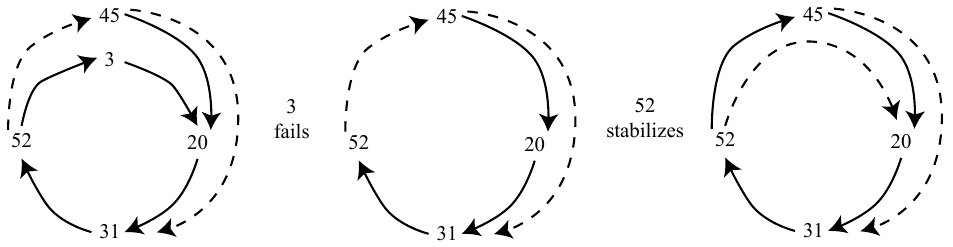}
\caption{A counterexample to a trial invariant.
Only the relevant pointers are drawn.}
\label{fig:wrap}
\end{figure*}

The {\it StabilizeFromPredecessor} step is simple.
The node queries its potential new successor for its successor list.
If the new successor is live, the node adopts it and its successor list.
If not, nothing changes.
Either way, the stabilization operation is complete.

\small
\begin{verbatim}
// StabilizeFromPredecessor step 

// newSucc value came from previous step
newSucc: Identifier;

query newSucc for newSucc.succList;
if (query returns before timeout)
   // new successor is live, adopt it
   succList = 
     append(newSucc,butLast(newSucc.succList));
   // else new successor is dead, no change
\end{verbatim}
\normalsize

At the completion of each stabilization operation, regardless of
the result, the stabilizing
node sends a message to its successor notifying the successor of its
presence as a predecessor.
On receiving this notification, a node executes a single-step
rectify operation, which may allow it to improve its predecessor
pointer.

\small
\begin{verbatim}
// Rectify step

// newPrdc value came from notification
newPrdc: Identifier;

if (between (prdc, newPrdc, myIdent)) 
   // newPrdc presumed live
   prdc = newPrdc;
else {
   query prdc to see if live;
   if (query returns before timeout) 
      no change;
   // live newPrdc better than dead old one
   else prdc = newPrdc;
};
\end{verbatim}
\normalsize

\section{Initialization and invariant}
\label{sec:initialization}

An {\it inductive invariant} is an invariant with the property that
if the system satisfies the invariant before any event,
then the system can be proved to satisfy the invariant after the 
event.
By induction, if the system's initial state satisfies the invariant,
then all system states satisfy the invariant.

Original Chord initializes a network with a single member that is its
own successor, {\it i.e.,} the initial network is a ring of size 1.
This is not correct, as shown in
Figure~\ref{fig:init} with successor lists of length 2.
Appendage nodes 62 and 37 start with both list entries equal to 48.
Then 48 fails, leaving members 62 and 37
with insufficient information to find each other.

Clearly the spirit of the operating assumption in 
Section~\ref{sec:overview} is that the chosen length of successor lists
should provide enough redundancy to ensure safe operation.
But we can hardly expect the successor lists to work if the redundancy
is thrown away by filling them with duplicate entries.
This is the problem with 
Figure~\ref{fig:init}---that 62 and 37 have no real redundancy in their
successor lists, so one failure disconnects them from the network.

For a member $n$ of a network with successor list length $r$ to enjoy
full redundancy, $n$ must have $r$ entries in its
successor list that are distinct from each other and from $n$.
For this to be possible, the network must have at least $r + 1$ members,
and the inductive invariant must imply that this is so.

The inductive invariant for Chord is the result of a very long and
arduous search, some of which is described in 
\cite{chord-arxiv}.
As an indication of the difficulty, consider Figure~\ref{fig:wrap},
which is a counterexample to a trial invariant consisting of the
conjunction of 
{\it AtLeastOneRing, AtMostOneRing, OrderedRing,
ConnectedAppendages, NoDuplicates,} and {\it OrderedSuccessorLists}.
Again $r = 2$.
Let an {\it extended successor list} be the concatenation of a node
with its successor list.
{\it NoDuplicates} has the obvious meaning that the entries in any
extended successor list are distinct.
{\it OrderedSuccessorLists} says that for any ordered sublist
{\it [x, y, z]}
drawn from a node's extended successor
list, whether the sublist is contiguous or not, 
{\it between [x, y, z]}
holds.

In Figure~\ref{fig:wrap},
the first stage satisfies the trial invariant, having duplicate-free
and ordered extended
successor lists such as {\it [52, 3, 45]} and {\it [45, 20, 31]}.
The appendage node 45 does not merge into the ring at the correct
place, but that is part of normal Chord operation (see \cite{chord-ccr}).
The second successor of ring node 52 points outside the ring, but
that is also part of normal Chord operation.
In the case shown in the figure, after two operations the ring has
become disordered.

Note that the four important properties
{\it AtLeastOneRing, AtMostOneRing, OrderedRing,} and
{\it ConnectedAppendages} are all stated in terms of which nodes are ring
members and which are appendage members.
Unfortunately, ring {\it versus} appendage
is not a stable characteristic of a member, but rather a fluid,
context-dependent property that changes easily.
In the figure, 45 changes from being an appendage
member to a ring member just because 3 fails.
In other examples, a ring member becomes an appendage member just because
a node fails.
So much of the difficulty of reasoning about Chord comes from the fact
that the obvious properties have no intrinsic stability or persistence.

The final inductive
invariant is much simpler than the earlier invariant used
in \cite{chord-arxiv}.
It also has the major advantage of not requiring an extra operating
assumption that is difficult to implement.

To explain the invariant, we must introduce the concept of a
{\it principal node}.
A principal node is a member that is not skipped by any member's
extended successor list.
For example, if 30 is a principal node, then
{\it [30, 34, 39]} and {\it [27, 30, 34]} and 
{\it [21, 27, 29]} can all be extended
successor lists, but {\it [27, 29, 34]} cannot be, because 30 is between
29 and 34, and would therefore be skipped.

The inductive invariant is the conjunction of only two properties,
{\it OneLiveSuccessor} and {\it SufficientPrincipals}.
{\it OneLiveSuccessor} simply says that every successor list has at
least one live entry.
{\it SufficientPrincipals} says that the number of principal members 
is greater than or equal to $r + 1$, where $r$ is the length of
successor lists.
A Chord network can be initialized in any state that satisfies the
invariant.

The proofs in Section~\ref{sec:idspace} will show that this deceptively
simple invariant implies all of
{\it AtLeastOneRing, AtMostOneRing, OrderedRing,
ConnectedAppendages, NoDuplicates,} and {\it OrderedSuccessorLists}.
Needless to say, it also implies that the network has a minimum size.
(Note that the first stage of 
Figure~\ref{fig:wrap} has no principal members, so the figure is
not a counterexample to the real invariant.)

A typical Chord network has $r$ from 2 to 5, so the
set of principals need only have 3 to 6 nodes.
Nevertheless, the existence of these few nodes protects the correctness
of a network with millions of members.
They wield great and mysterious powers!

\section{Comparison of the versions}
\label{sec:diff}

In the original version of Chord,
the {\it join, stabilize,} and {\it notified}
operations are defined as pseudocode in
\cite{chord-sigcomm} and \cite{chord-ton}, as is the initialization.
These papers do not provide details about failure recovery,
so the definition of the original version of Chord is completed
by adding the pseudocode for
{\it reconcile, update,} and {\it flush} operations from 
\cite{chord-podc}.
The ``new'' version of Chord is the one specified in this paper.
The following table shows how operations of the two versions correspond.
Although {\it rectify} in the new version is similar to
{\it notified} in the original version, it seems more consistent to use
an active verb form for its name.

\begin{center}
\footnotesize
\begin{tabular}{|c|c|}
\hline
{\bf original} & {\bf new} \\ \hline
join +	& join \\
reconcile &  \\ \hline
stabilize + & stabilize \\
reconcile + & \\
update & \\ \hline
notified + & rectify \\
flush & \\ \hline
\end{tabular}
\normalsize
\end{center}

In both original and new versions of Chord, members schedule their own
maintenance operations except for {\it notified} and
{\it rectify}, which occur when a member is notified by a
predecessor.
Although the operations
are loosely expected to be periodic, scheduling is not formally
constrained.
As can be seen from the table, 
multiple smaller operations from the old version
are assembled into larger new operations.
This ensures that the successor lists of members are always fully
populated with $r$ entries,
rather than having missing entries to be filled in by later
operations.
An incompletely populated successor list might lose (to failure)
its last live
successor.
If the successor list belongs to an appendage member, this would mean
that the appendage can no longer reach the ring, which is
a violation of {\it ConnectedAppendages} \cite{chord-ccr}.

Another systematic change from the old version to the new is that,
before incorporating a pointer to a node into its state, a
member checks that the node pointed to is live.
This prevents cases where a member replaces a pointer to a live node with
a pointer to a dead one.
A bad replacement can also cause a successor list to have no live
successor.
If the successor list belongs to a ring member, this will cause a
break in the ring, and a 
violation of {\it AtLeastOneRing}.
Together these two systematic changes also prevent
scenarios in which the ring becomes disordered or breaks into
two rings of equal size
(violating {\it OrderedRing} or {\it AtMostOneRing}, respectively
\cite{chord-ccr}).

A third systematic change was necessary because the original version
does not say anything precise about communication between nodes,
and does not say anything at all about atomic steps and maintaining
a shared-state abstraction.
The new operations are specified in terms of atomic steps,
and the rules for maintaining a shared-state abstraction are stated
explicitly.

The other major difference is the initialization, as discussed in
Section~\ref{sec:initialization}.

In addition to these systematic changes, a number of small changes
were made.
Some were due to problems detected by Alloy modeling and analysis
of the original version.
Others were required to ensure that, after each atomic step of a
stabilize operation, the global state satisfies the invariant.

These differences do not change the efficiency of Chord operations in any
significant way.
Checking some pointers to make sure they point to live nodes (new version)
requires more queries than in the original version.
On the other hand, in the original version {\it stabilize, reconcile,} and
{\it update} operations are all separate, and can all entail queries.
In this respect the original version requires more queries than the new 
version.

There is an additional bonus in the new version for implementers.
Consider what happens when a
member node fails, recovers, and
wishes to rejoin, all of which could occur within a short period of
time.
It was previously thought necessary
for the node to wait until all
previous references to its identifier had been 
cleared away (with high probability), 
because obsolete pointers could make the state incorrect. 
This wait was included in the first Chord
implementation \cite{excuses}.
Yet the wait is unnecessary, as Chord is provably
correct even with obsolete pointers.

In the spirit of \cite{sitmorris}, it is a good security practice
to monitor that invariants are satisfied.
Both the conjuncts of the inductive invariant are global, and thus
unsuitable for local monitoring.
The right properties to monitor are {\it NoDuplicates} and
{\it OrderedSuccessorLists}, which
can be checked on individual successor lists.
These are properties that must be true for Chord networks of
any size.

Although the new initialization with $r + 1$ principal nodes
may not be inefficient, it is certainly more difficult to
implement than initialization of original Chord.
An alternative approach might be to start the network with a single
node, and monitor the network as a whole
until it has $r + 1$ principal nodes.
For example, all nodes might send their successor lists (whenever there
is a change) around the ring, to be collected and checked by the single
initial node.
Once the initial node sees a sufficient set of principal
nodes, it could send a signal around the ring that monitoring is no
longer necessary.
This scheme is discussed further in Section~\ref{sec:preservingbase}.

\section{Reasoning about ring structures in identifier spaces}
\label{sec:idspace}

\subsection{Theorems about identifier spaces}

An identifier space is built from a finite, totally-ordered (in the
usual binary sense) set.
An identifier space also has a total ternary order 
{\it between,}
defined in Alloy as:
\small
\begin{verbatim}
pred between[n1,nb,n2: Node] {
   lt[n1,n2] =>   ( lt[n1,nb] && lt[nb,n2] )
             else ( lt[n1,nb] || lt[nb,n2] ) }
\end{verbatim}
\normalsize
where
\small
{\tt lt, \&\&, ||}
\normalsize
are the notations for less than (in the total binary order),
logical and, and logical or,
respectively.
The definition has the form of an if-then-else expression.
This definition
has the same semantics as the pseudocode predicate {\it between}
in Section~\ref{sec:state}.

Informally, order in the identifier space ``wraps around'' from
the last element of the binary order to the first.
Because of this wraparound, two elements cannot be compared,
which is why order in an identifier space must be ternary.

In this section definitions and theorems about identifier spaces
are presented in the Alloy syntax.
In the Alloy model the concepts of identifier and node (potential
network member) are conflated, so that {\tt Node} is declared as a
totally ordered set upon which an identifier space is built.
Details about the Alloy model and bounded verification can be
found in Section~\ref{sec:alloy}.
These theorems have been proven for unbounded identifier spaces
using merely substitution and simplification.

Here is a simple theorem in Alloy syntax:
\small
\begin{verbatim}
assert AnyBetweenAny {
   all disj n1,n2: Node | between[n1,n2,n1] }
\end{verbatim}
\normalsize
{\it AnyBetweenAny}
says that for any distinct (disjoint)
{\it n1} and {\it n2, n2} is between {\it n1} and {\it n1}.

For proofs, we also need a different predicate {\it includedIn},
which is like {\it between} except that the included
identifier can be equal to either of the boundary identifiers:
\small
\begin{verbatim}
pred includedIn[n1,nb,n2: Node] {
  lt[n1,n2] =>   ( lte[n1,nb] && lte[nb,n2] )
            else ( lte[n1,nb] || lte[nb,n2] ) }
\end{verbatim}
\normalsize
In the {\it AnyIncludedInAny} theorem,
the two arguments need not be disjoint:
\small
\begin{verbatim}
assert AnyIncludedInAny {
   all n1,n2: Node | includedIn[n1,n2,n1] }
\end{verbatim}
\normalsize

A very useful theorem allows us to reason about the fact or assumption
that 
{\it between} 
does {\it not} hold.
\small
\begin{verbatim}
assert IncludedReversesBetween { 
   all disj n1,n2: Node, nb: Node |
       ! between[n1,nb,n2] 
   <=> includedIn[n2,nb,n1]       }
\end{verbatim}
\normalsize
Provided that the boundaries of an interval are distinct, if an
identifier {\it nb} 
cannot be found in the portion of the identifier
space from {\it n1} to {\it n2} (exclusive),
then it must be found in the portion of the identifier space
from {\it n2} to {\it n1} (inclusive).

The viewpoint of this paper is that identifier spaces have less
structure than algebraic rings.
Algebraic rings are generalizations of integer arithmetic, with operators
such as sum and product that combine quantities.
In Chord identifiers are not quantities, and it makes no sense to
add or multiply them.
This is in contrast to the formalization of Pastry 
\cite{pastry-proof}, where distance in the identifier space is assumed
to be meaningful and is used in the protocol.

\subsection{Theorems about successor lists}

This section introduces definitions and theorems about ring-shaped
networks whose structure is based on successor lists
in an identifier space.
A number of terms concerning successor lists in network
states were introduced briefly in 
Section~\ref{sec:initialization}.
For clarity, they will be redefined here.

\vspace{.1 in}
{\it Definition:} An {\it extended successor list} (ESL) 
is a successor list
with the node that owns it prepended to the list.
The length of an ESL is $r + 1$.

\vspace{.1 in}
{\it Definition:} A {\it principal node} is a member that is not
skipped by any ESL.  
That is, for all principal nodes {\it p}, there is no
contiguous pair {\it [x, y]} in any ESL such that {\it between [x, p, y]}.

\vspace{.1 in}
{\it Definition:} The property {\it OneLiveSuccessor} holds in a
state if every member has at least one live entry in its successor list.

\vspace{.1 in}
{\it Definition:} The property {\it SufficientPrincipals} holds in a
state if the number of principal nodes is greater than or equal to
$r + 1$.

\vspace{.1 in}
{\it Definition:} The property {\it Invariant} is the conjunction
of {\it OneLiveSuccessor} and {\it SufficientPrincipals}.

\vspace{.1 in}
{\it Definition:} The property {\it NoDuplicates} holds in a
state if every ESL has $r + 1$ distinct entries.

\vspace{.1 in}
{\it Definition:} The property {\it OrderedSuccessorLists} holds in a
state if for all sublists {\it [x, y, z]} of ESLs, whether contiguous
sublists or not, {\it between [x, y, z]}.
\vspace{.1 in}

The remainder of this section proves that {\it Invariant} implies
the successor-list properties {\it NoDuplicates} and
{\it OrderedSuccessorLists}.

\vspace{.1 in}
{\it Theorem:} In any ring structure whose state is maintained in
successor lists, {\it Invariant} implies {\it NoDuplicates}.

{\it Proof:}

Contrary to the theorem, assume that there is a network state for which
{\it Invariant} is true and {\it NoDuplicates} is false.
Then some node has an extended successor list with the form
{\it [ ..., x, ..., x, ... ]} for some identifier {\it x}.

From {\it AnyBetweenAny}, 
for all principal nodes {\it p} 
distinct from  {\it x, between [x, p, x]}.
Because of the definition of principal nodes, all of the principal
nodes distinct from {\it x} must be listed in the ellipsis
between the two occurrences of {\it x} in the successor list.

From {\it SufficientPrincipals}, the portion of the extended successor list
{\it [x, ..., x]} must have length at least $r + 2$, because there are at
least $r$ principal nodes distinct from {\it x}.
But the length of the entire extended successor list is $r + 1$, which
yields a contradiction.
$\Box$

\vspace{.1 in}
{\it Theorem:} In any ring structure whose state is maintained in
successor lists, {\it Invariant} implies {\it OrderedSuccessorLists}.

{\it Proof:}

Contrary to the theorem, assume that there is a network state for which
{\it Invariant} is true and {\it OrderedSuccessorLists} is false.
Then some node has an ESL with the form
{\it [ ..., x, ..., y, ..., z, ... ]} where not {\it between [x, y, z]}.
From the previous theorem, {\it x, y,} and {\it z} are all distinct.

\begin{figure}
\centering
\includegraphics[scale=0.80]{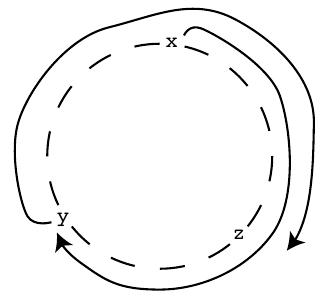}
\caption{The dashed line depicts the identifier space.
The solid arrows show the path around the identifier space
of a segment of an ESL {\it [x, ..., y, ..., z].}}
\label{fig:osl}
\end{figure}

From {\it IncludedReversesBetween}, {\it includedIn [z, y, x]}.
If we visualize an identifier space as a ring ordered clockwise
(as in Figure~\ref{fig:osl}),
the disordered ESL segment {\it [x, ..., y, ..., z]} wraps around the
identifier ring
touching the identifier space first at {\it x},
passing by {\it z}, touching at {\it y}, passing by {\it x} again,
then finally touching at {\it z}.

It is easy to see that the only identifier in the entire identifier
space that is not skipped by this ESL is $y$.
Yet there must be more than one principal node, 
which is a contradiction.
$\Box$
\vspace{.1 in}

\begin{figure*}
\centering
\includegraphics[scale=0.80]{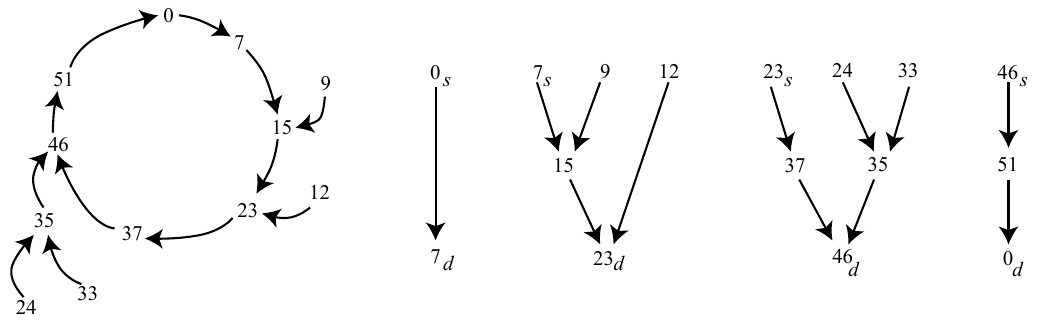}
\caption{For a network, the {\it bestSucc} relation is pictured on the
left, and the {\it splitBestSucc} relation is pictured on the right.
Although it cannot be seen from best successors only, 
the principal nodes are 0, 7, 23, and 46.}
\label{fig:splitbestsucc}
\end{figure*}

\subsection{Theorem about networks built on successor lists}
\label{sec:theorem3}

This section is concerned with proving that {\it Invariant} implies
the four necessary properties introduced in Section~\ref{sec:overview}.

\vspace{.1 in}
{\it Definition:} A network member's {\it best successor}
is the first live node in its successor list.

\vspace{.1 in}
{\it Definition:} A {\it ring member} is a network member that can
be reached by following the chain of best successors beginning at
itself.

\vspace{.1 in}
{\it Definition:} An {\it appendage member} is a network member that is
not a ring member.

\vspace{.1 in}
{\it Definition:} The property {\it AtLeastOneRing} holds in a
state if there is at least one ring member.

\vspace{.1 in}
{\it Definition:} The property {\it AtMostOneRing} holds in a
state if, from every ring member, it is possible to reach every other
ring member by following the chain of best successors beginning at
itself. 

\vspace{.1 in}
{\it Definition:} The property {\it OrderedRing} holds in a
state if on the unique ring, the nodes are in identifier order.
That is, if nodes 
{\it n1} and {\it n2} are ring members,
and {\it n2} is the best successor of {\it n1}, 
then there is no other ring member 
{\it nb} 
such that 
{\it between [n1, nb, n2]}.

\vspace{.1 in}
{\it Definition:} The property {\it ConnectedAppendages} holds in a
state if, from every appendage member, a ring member can be reached
by following the chain of best successors beginning at itself.

\vspace{.1 in}
{\it Theorem:} In any ring structure whose state is maintained in
successor lists, {\it Invariant} implies 
{\it AtLeastOneRing},
{\it AtMostOneRing},
{\it OrderedRing}, and
{\it ConnectedAppendages}.

{\it Proof:}

The best-successor relation {\it bestSucc} is a binary relation on
network members.
We define from it a relation {\it splitBestSucc} that is the same except
that every principal node {\it p} is replaced by two nodes
$p_s$ and $p_d$, where $p_s$ ({\it s} for source) is in the domain of
the relation but not the range,
and $p_d$ ({\it d} for destination) is in the range of
the relation but not the domain.
Figure~\ref{fig:splitbestsucc} displays as graphs the
{\it bestSucc} and {\it splitBestSucc} relations
for the same network. 
It is possible to deduce many properties of the {\it splitBestSucc} graph,
as follows:

(1) 
From {\it Invariant}, every member has a best successor.
So the only nodes with no outgoing edges are $p_d$ nodes representing
principal members only as {\it being} best successors.

(2) $p_s$ nodes have no incoming edges, as they represent principal
nodes only as {\it having} best successors.
There can be other nodes with no incoming edges, because there can be
members that are no member's successor.

{\it Note:} The next few points concern maximal 
paths in the {\it splitBestSucc} graph.
These are paths beginning at
nodes with no incoming edges.
By definition, they can only end at $p_d$ nodes, and can have no
internal nodes representing principal nodes.

(3)
Just as a successor list does not skip principal nodes, a maximal path of
best successors does not skip principal nodes.
That is because an adjacent pair {\it [x, y]} in a chain of
best successors is taken from the successor list of {\it x},
and if there are any entries in the successor list between
{\it x} and {\it y}, they are dead.

(4) A maximal path is acyclic.
Contrary to this statement, assume that the path contains a cycle
{\it x leads to x}.
By definition, the path has no nodes representing principal nodes.
Yet the path traverses the entire identifier space, so it skips all
principal nodes, which contradicts (3).

From (1-4), we know that the graph of {\it splitBestSucc} is an
inverted forest (a ``biological'' forest, with roots on the bottom and
leaves on the top).
Each tree is rooted at a $p_d$ node.

(5) A maximal path is ordered by identifiers.
Contrary to this statement, let the path contain 
{\it [x, ..., y, ..., z]} where not {\it between [x, y, z]}.
Because the path is acyclic,
{\it x, y,} and {\it z} are all distinct.
From {\it IncludedReversesBetween}, {\it includedIn [z, y, x]}.
So the disordered path segment {\it [x, ..., y, ..., z]} wraps around the
identifier ring exactly as the ESL does in Figure~\ref{fig:osl}.

As in the proof of {\it OrderedSuccessorLists}, this segment skips
every identifier in the entire identifier space except $y$.
There must be more than one principal node, so this segment skips
a principal node, contradicting (3).

{\it Note:} The next two points concern the mapping from
$p_s$ nodes to $p_d$ nodes derived from maximal paths in
{\it splitBestSucc}.
They show that it is a bijection, {\it i.e.,} one-to-one and onto.

(6)
Every $p_s$ node is a leaf of exactly one tree.
It must be a leaf of some tree, because it begins a path of best
successors that must end at a $p_d$ node.
It cannot be a leaf of more than one tree, because no node has more than
one best successor.

(7) Every tree rooted at a $p_d$ node has at least one leaf $p_s$,
which is the principal node closest to $p_d$ in reverse identifier
order.
It cannot have two such leaves $p1_s$ and $p2_s$, because the source
principal closer to the destination principal would be skipped by the
path of the farther source principal.

{\it Summary:}

In terms of {\it splitBestSucc}, 
a ring is formed by the concatenation of the unique maximal paths,
one from each tree in the forest, starting at $p_s$ nodes.
This proves {\it AtLeastOneRing}.
The relation cannot separate into two rings, because each of them
would have a maximal path that skips a principal node in the other;
this proves {\it AtMostOneRing}.
From (5) we know that the ring is ordered by identifiers,
so {\it OrderedRing} holds.
All the nodes not on these unique maximal paths are appendage members,
and each has a path to a principal node on the ring, so
{\it ConnectedAppendages} holds.
$\Box$
\vspace{.1 in}

\section{Proof of Chord correctness}
\label{sec:proof}

This section presents the proof of the theorem given in 
Section~\ref{sec:overview}:

\vspace{.1 in}
{\it Theorem:} In any execution state, if there are
no subsequent join or fail operations, then eventually the network will
become Ideal and remain Ideal.
\vspace{.1 in}

The most important part of this theorem is knowing that {\it Invariant}
holds in all states, because this property and the properties it 
implies are the ones that all Chord users can count on at all times.
We do not expect churn (joins and failures) to ever stop long enough
for a network to become {\it Ideal.}
Rather, this part of the theorem simply tells us that the repair
algorithm always makes progress, and cannot get into unproductive loops.

\subsection{Establishing the invariant}
\label{sec:invariant}

First it is necessary to prove that {\it Invariant,} which is
true of any initial state (as specified in 
Section~\ref{sec:initialization}), is preserved by every atomic step of the
protocol.

We begin with a failure step, because it requires a constraint
based on the operating assumption in Section~\ref{sec:overview}:
a member cannot fail if it would leave another member with no live
successor.
In other words, failure steps preserve the property of 
{\it OneLiveSuccessor} by operating assumption.
No other kind of step can violate
{\it OneLiveSuccessor}.

The other conjunct of {\it Invariant} is
{\it SufficientPrincipals}, which says that the number of principal nodes
must be at least $r + 1$.
Rectify operations cannot violate this property, as they do not affect
successor lists.
In this section we will show that failure steps of non-principal
nodes, join steps,
{\it StabilizeFromSuccessor} steps, and {\it StabilizeFromPredecessor}
steps do not cause principal nodes to become skipped in successor lists.
This is the only way that they could violate {\it SufficientPrincipals.}
The remaining case, that of failures of principal nodes, will be
discussed in the next section.

Failure of a non-principal member {\it n} causes the disappearance of 
{\it n}'s successor list.
But only being skipped in a successor list can make a node non-principal,
so the disappearance of {\it n}'s successor list cannot make another
node non-principal.

In a successful join, the new ESL created is
{\it [myIdent, newPrdc.succList]}.
We know that there is no principal node between 
{\it myIdent} and {\it head (newPrdc.succList)}, because 
at the time of the query there is no principal node between
{\it newPrdc} and {\it head (newPrdc.succList)}, and
{\it myIdent} is between those two.
We also know that {\it newPrdc.succList} cannot skip a principal node,
by definition.

There are two cases in a {\it StabilizeFromSuccessor} step where
a successor list is altered.
In the first case the new ESL is a concatenation of pieces of the
ESLs of the stabilizing node and its first successor, joined where they
overlap at the first successor.
Since neither of the original ESLs can skip a principal node, their
overlap cannot, either.

In the second case a dead entry is removed from the stabilizing node's
list, which cannot cause it to skip a principal.
This leaves an empty space at the end which is temporarily padded with
the last real entry plus one.
This is the only value choice that preserves the invariant by 
guaranteeing that no principal node is skipped by accident.
It does not matter whether the artificial entry points to a real node
or not, as it will be gone by the time that the stabilization operation
is complete.

There is only one case in a {\it StabilizeFromPredecessor} step where
a successor list is altered.
The new ESL created is
{\it [myIdent, newSucc, butLast (newSucc.succList)]}.
In the previous {\it StabilizeFromSuccessor} step, this node tested
that {\it between [myIdent, newSucc, head (succList)]}.
This node cannot make any other changes to its successor list
between that step and this {\it StabilizeFromPredecessor} step,
so it is still true.
Therefore we
know that there is no principal node between 
{\it myIdent} and {\it newSucc}, because 
there is no principal node between
{\it myIdent} and {\it head (succList)}, and
{\it newSucc} is between those two.
We also know that {\it [newSucc, butLast (newSucc.succList)]} 
cannot skip a principal node, because it is part of the ESL of
{\it newSucc}.

\subsection{Failure of principal nodes}
\label{sec:preservingbase}

The fundamental reason why a Chord network must have $r + 1$ principal
nodes is the need to prove {\it NoDuplicates}.
Without {\it NoDuplicates} we cannot justify the operating assumption
that a member always has a live successor, because the assumption is
based on the full redundancy provided by ESLs with $r + 1$ distinct
entries.

Apart from initialization, a member of a Chord network becomes a
principal node when it has been a member long enough so that every
node that should know about it does know about it.
More specifically, it should appear in the successor lists of its
$r$ predecessors, which will happen after a sequence of $r$ stabilizations
in which each predecessor learns about the node from its successor.

It is extremely important that Section~\ref{sec:invariant} showed that 
none of the operations or steps of operations discussed there can
demote a node from principal to non-principal.
In other words, the {\it only} action that can reduce the size of the
set of principal nodes is failure of a principal node itself.

As a Chord network grows and matures, a significant fraction of its
nodes will be members long enough to become principals.
This means that the number of principal nodes is proportional to the
size of the network;
once the network is large enough there is no possibility that
{\it SufficientPrincipals} will be violated.
Section~\ref{sec:diff} presented the idea of global monitoring of
small Chord networks as a way to implement initialization with 
$r + 1$ principal nodes.
It is a simple change to continue monitoring until the number of
principal nodes has reached some multiple of $r$, after which the
network is safe.

\subsection{Queries have no circular waits}

Section~\ref{sec:shared} explained how inter-node queries must be
organized to maintain a shared-state abstraction.
Sometimes a node must delay answering a query because it is waiting
for the answer to its own query, which raises the specter of deadlock
due to circular waiting.

Note that a rectify step only queries to see if a node is still alive,
and does not read any of the node's state.
Queries like these can always be answered immediately, so cannot
cause waiting.

Note also that a join step requires a query, but no other node can be
querying a node that has not joined yet.
So the joining node, also, cannot be part of a circular wait.

This leaves queries due to the two stabilization steps, which are
always directed to first successors or potential first successors.
This means that, if there is a circular wait due to queries, it must
encompass the entire ring.
This possibility is sufficiently remote to ignore.\footnote{The
formal model uses shared-memory communication as an abstraction
of queries. 
Waiting is not modeled, so this case is not a problem for formal
analysis.}

\subsection{Proving progress}
\label{sec:progress}

This section shows that in a network satisfying {\it Invariant},
if there are no join or fail operations, then eventually the network will
become {\it Ideal}
and remain {\it Ideal} (as defined in Section~\ref{sec:overview}).

Progress proceeds in a sequence of phases.
In the first phase, all leading dead entries are removed from
successor lists, so that every member's successor is its best
successor.
Every time a member with a leading dead entry begins stabilization,
it first executes a {\it StabilizeFromSuccessor} step, which will
remove the leading dead entry.
It will continue executing {\it StabilizeFromSuccessor} steps until
all the leading dead entries are removed.
Eventually all members will stabilize (this is an operating assumption),
after which all leading dead entries will be removed from all
successor lists.

Needless to say, these effective 
{\it StabilizeFromSuccessor} steps can be interleaved with other 
stabilize and rectify operations.
However, rectify operations do not change successor lists.
Even if a stabilization operation causes a node to change its successor,
the steps are carefully designed so that the node will not change its
successor to a dead entry.
So, in the absence of failures, eventually all successors will be
best successors, and will remain so.

In the second phase, which can proceed concurrently with or subsequent to
the first phase, all successors and predecessors become correct.
Let $s$ be the current size of the network (number of members).
This number is only changed by join and fail operations, and not by
repair operations, so it remains the same throughout a repair-only
phase as hypothesized by the theorem.
The error of a successor or predecessor
is defined as 0 if it points to the
first successor (respectively, predecessor) in identifier order,
1 if it
points to the second successor (predecessor) in
identifier order, . . . $s - 1$ if it points to
the least correct member, and
$s$ if it points to a dead node.

Whenever there is a merge in the {\it bestSucc} or
{\it splitBestSucc} graph
(see Figure~\ref{fig:splitbestsucc}),
there are two nodes {\it n1} and 
{\it n2} with successors merging at {\it n3}, and for some choice of
symbolic names, {\it between [n1, n2, n3]}.
There are three cases: 
(1) {\it n3.prdc} (the current predecessor of {\it n3})
is better (has a smaller error) than {\it n2}, meaning that
{\it between [n2, n3.prdc, n3]};
(2) {\it n3.prdc} is {\it n2};
(3) {\it n3.prdc} is worse (has a larger error) than {\it n2},
meaning that
{\it between [n3.prdc, n2, n3]}.
In each of these three cases there is a sequence of enabled operations
that will reduce the cumulative error in the network, as follows:

\vspace{.1 in}
{\it Case 1:}
Either {\it n1} or {\it n2} stabilizes, adopting {\it n3.prdc} as
its successor and reducing the error of its successor.
When the stabilizing node notifies {\it n3.prdc} and 
{\it n3.prdc} rectifies, it will change its predecessor pointer
if and only if the change reduces error.

\vspace{.1 in}
{\it Case 2:}
{\it n1} stabilizes, adopting {\it n2} as
its successor and reducing the error of its successor.
When {\it n1} notifies {\it n2} and 
{\it n2} rectifies, it will change its predecessor pointer
if and only if the change reduces error.

\vspace{.1 in}
{\it Case 3:}
{\it n2} stabilizes, which will not change its successor,
but will have the effect of notifying {\it n3}.
When {\it n3} rectifies, it will reduce the error of its predecessor
by changing it to {\it n2}.
\vspace{.1 in}

These cases show that, as long as there is a merge in the {\it bestSucc}
graph, some operation or operations are enabled that will reduce the
cumulative error of successor and predecessor pointers.
Equally important, all operations are carefully designed so that a
change never increases the error.
At the same time, some of these operations will reduce the number
of merges.
For example, in Figure~\ref{fig:splitbestsucc}, let the merge of
24 and 33 at 35 be an example of Case 2.
When 24 changes its successor to 33, which is not currently the
successor of any node, the total number of merges is reduced.

As the network is finite, eventually 
there will be no merges in the
{\it bestSucc} graph, which means that every node is a ring member.
Because the ring is always ordered, the errors of all successors will
be 0.
The errors of all predecessors will also be 0, because whenever a
successor pointer reaches its final value by stabilization, it
notifies its successor.
That node will update its predecessor pointer, and will never again
change it, because no other candidate value can be superior.
This is the completion of the second phase.

In the third and final phase, after all successors are correct,
the tails of all successor lists become
correct (if they are not already).
Let the error of a successor list tail of length $r - 1$ be defined as the
length of its suffix that does not match
its member's successor's successor list.

Let {\it n2} be the successor of {\it n1}, and let the error of
{\it n2}'s successor list be $e$.
When {\it n1} stabilizes, the error of its successor list becomes
{\it max}$(e - 1,0)$, 
as it is adopting {\it n2}'s successor list, after first
prepending a correct entry ({\it n2}) and dropping an entry at the end.
Thus improvements to successor lists propagate backward in identifier
order.
In the worst case, after a backward chain of $r - 1$ stabilizations,
the successor list of the last node of the chain will be globally correct.
The correct list will continue propagating backward, leaving
correctness in its wake.
$\Box$

\section{The Alloy model and bounded verification}
\label{sec:alloy}

As introduced in Section~\ref{sec:intro}, there is an Alloy
model 
including specification of the operations, correctness properties, and 
assertions of the 
proof.\footnote{\url{http://www.research.att.com/~pamela} $>$ How to
Make Chord Correct.}
The reasons for using Alloy for this purpose can be found in
\cite{compare}.

The Alloy proof is direct rather than insightful.
For example, there are assertions of all the theorems in 
Section~\ref{sec:idspace}.
The Alloy Analyzer uses exhaustive enumeration to verify
automatically that the theorems are true for all model instances up
to some size bounds (see below).
But unlike Section~\ref{sec:idspace}, this verification gives no
insight into why the theorems are true.

The Alloy proof treats progress somewhat differently from 
Section~\ref{sec:progress}.
The model defines enabling predicates for all operation cases, 
where an enabling predicate is true if and only if a step or
sequence of steps is enabled and will change the state of the network
if it occurs.
An assertion states that if a network is not {\it Ideal,} some operation
is enabled that will change the state.
Another assertion states that if a network is {\it Ideal,} no operation
will change the state.

What is missing from the formal treatment of progress is the argument
that every change makes progress.
This is provided by the error metrics in the informal proof.
In principle the error metrics could be defined and checked in Alloy,
but experience suggests that this would be awkward and computationally
complex.

The model is and has been an indispensable part of this research,
for two reasons:
First, it protects against human error in the long informal proof.
Second, it was a necessary tool for getting to the proof.
Without long periods of model exploration, it would not have been
possible to discover that the obvious invariants are not sufficient,
nor to discover an invariant that is.
Without the formal model and automated verification, one wastes too
much time trying to prove assertions that are not true.

The model is analyzed for all instances with $r \leq 3$ and $n \leq 9$,
where $n$ is the size of the identifier/node space.
For the largest instances,
the possible number of nodes is more than twice the
sufficient number of principal nodes.

It is worth noting what experimenting with models and bounds is like.
With $r = 2$, many new counterexamples (to the current draft model)
were found by increasing the number of nodes from 5 to 6, and no new
counterexamples were ever found by increasing the number of nodes
from 6 to 7 or more.
No new counterexamples were ever found by increasing $r$ from 2 to 3.
This makes $r = 3$ and $n = 9$ seem more than adequate.

\section{Related and future work}
\label{sec:related}

Other researchers have found problems with Chord implementations.
Freedman {\it et al.} found that the assumption of bidirectional
network communication can be violated in practice \cite{chord-nontrans}.
Model-checking of code has been used to find bugs in
implementations of Chord \cite{mace,crystalball}.
No previous work except \cite{chord-ccr}, however, has discovered any
problems with the specification of Chord.

Although other researchers have verified properties of DHTs
\cite{chord-sweden,pastry-proof}, they have not considered failures,
which are by far the most difficult part of the problem.
Other work on verifiable ring maintenance operations \cite{ringtop}
uses multi-node atomic operations, which are avoided by Chord.
These studies use a variety of techniques.
Adhering to the terminology used so far, in which ``informal''
means rigorous but not machine-checked, and ``formal'' means
machine-checked:  \cite{ringtop} employs
informal specification and proof, while
\cite{chord-sweden} has formalizations of a specification and an
implementation in $\pi$-calculus, plus an informal proof of bisimulation.
Only \cite{pastry-proof} is completely formal, with specification 
and proof in TLA.

The Alloy-only proof in an earlier version of this work
\cite{chord-arxiv} has been used as a test case for the Ivy proof
system \cite{ivy}.
The Ivy version of Chord is a significant simplification, as it has
larger atomic operations, and limits the length of successor lists to 2.
Most importantly, it assumes the property {\it OneLiveSuccessor} without 
maintaining the {\it NoDuplicates} property that justifies the
assumption (see Section~\ref{sec:preservingbase}).
Nevertheless, the study yielded promising results with respect to its
goal of automatically generating invariants.

Now that our understanding of the protocol has a firm foundation,
it should be possible to exploit this knowledge to improve
peer-to-peer networks further.
If these efficient networks become more robust, they may find a whole
new generation of applications.

For example,
the assumptions of reliable network communication, bidirectional
network communication, and perfect detection of failures through
timeouts are all related and all suspect.
With a bit more overhead, it
might be possible to weaken these assumptions without compromising
Chord's modest invariant.

It is certainly possible to enhance security just by checking local
invariants, 
and it may be possible to improve 
enhancements such as
protection against malicious peers
\cite{awerbuch-robust,chord-byz,sechord},
key consistency and data consistency \cite{scatter},
range queries \cite{rangequeries},
and atomic access to replicated data \cite{atomicchord,etna}.
The first step is to update this work with the new correct
specification, then revisit the possible improvements in light of the
new invariant.

\section{Conclusion}

The Chord ring-maintenance protocol is interesting in several ways.
The design is 
extraordinary in its achievement of consistency and
fault-tolerance with such simplicity,
so little synchronization overhead, and such a weak assumption
about the occurrence of failures.
Unlike most protocols, which work according to self-evident
principles, it is quite difficult to understand how and why Chord works.

As a case study in practical verification, the Chord project illustrates
the value of a variety of techniques.
Simple analysis for bug-finding \cite{chord-ccr}, 
fully automated verification through bounded model-checking 
\cite{chord-arxiv},
and informal mathematical proof, all had important roles to play.

\section*{Acknowledgments}

Helpful discussions with
Bharath Balasubramanian,
Ernie Cohen,
Patrick Cousot,
Gerard Holzmann,
Daniel Jackson,
Arvind Krishnamurthy,
Leslie Lamport,
Gary Leavens,
Pete Manolios,
Annabelle McIver,
Ken McMillan,
Jay Misra,
Andreas Podelski,
Emina Torlak,
Natarajan Shankar, and
Jim Woodcock
have contributed greatly to this work.
The anonymous reviewers significantly improved its presentation.

\bibliographystyle{IEEEtran}
\bibliography{proved}

\end{document}